\documentclass[12pt]{iopart}
\usepackage{graphicx}
\usepackage{iopams}

\begin{document}
\title[Kinematics of accelerated systems]{A new relativistic kinematics of  accelerated systems}
\author{Yaakov Friedman and
Yuriy Gofman}
  \address{Jerusalem College of
Technology\\P.O.B. 16031 Jerusalem 91160, Israel\\email: friedman@jct.ac.il\\
Published: Physica Scripta 82 (2010) 015004 }

\begin{abstract}
We consider transformations between uniformly accelerated systems, assuming that the Clock Hypothesis is false.
We use the proper velocity-time description of events rather than the usual space-time description
in order to obtain linear transformations. Based on the generalized principle of relativity and the ensuing symmetry, we obtain transformations of Lorentz-type. We predict the existence of a maximal acceleration and
time dilation due to acceleration. We also predict a Doppler
shift due to acceleration of the source in addition to the shift due to the source's velocity.
Based on our results, we explain the  W. K\"{u}ndig experiment, as reanalyzed  by Kholmetski \textit{et al}, and obtain an estimate of the maximal acceleration.


\end{abstract}

 \maketitle

\section{Introduction}

Transformations between uniformly accelerated systems in flat space-time may provide a
connection between special and general relativity.
In order to study accelerated systems, A. Einstein introduced the Clock Hypothesis, which states that the ``rate of an accelerated clock is identical to that
of the instantaneously comoving inertial clock."  Not all physicists agree with this
hypothesis. L. Brillouin (\cite{Brilluin} p.66) wrote that ``we do not know and should not
guess what may happen to an accelerated clock." If we assume the validity of the Clock hypothesis, then the space-time transformation between
 accelerated systems are well known, see \cite{Moler}
 and others.
 Here we present a systematic approach for transformations between
 accelerated systems \textit{without} assuming the Clock Hypothesis.
Our approach to describing transformations between two uniformly accelerated systems is based
on the symmetry following from  the general principle of
relativity.

To simplify our derivations, we will consider a one dimensional space.
To reach our conclusions, it is enough to consider this simplified case.
We will clarify in Section 2 the precise meaning of
uniform acceleration and the notion of a system uniformly accelerated with
respect to an inertial system. It is clear that in order to
describe transformations between two systems which are uniformly
accelerated with respect to an inertial system, it is enough to
describe the transformation from an inertial system to a system
uniformly accelerated with respect to this system. We will
decompose this transformation into a product of a transformation
between an inertial system and a system comoving with a uniformly
accelerated system and a transformation from the comoving system
to the uniformly accelerated system. For the first transformation,
explicit space-time transformations are known.

For the second transformation, we will consider more general
transformations between two comoving systems uniformly accelerated
with respect to an inertial system. The method of solving this
problem is based on the method used in \cite{FG02/2} and, in more detail,
in \cite{F04}, for deriving the
Lorentz transformation between inertial systems from the principle of special
relativity and the ensuring symmetry. In Section 3, we introduce
a new proper velocity-time description of events, replacing the usual
space-time description. This will make the transformations linear.

In Section 4, we derive general proper velocity-time transformations between
comoving uniformly accelerated systems. The derivation is based on
the General Principle of Relativity and the ensuring symmetry. By careful
choice of reference frames, we derive linear Lorentz-type
transformations which depend on a constant $\kappa$. If the Clock
Hypothesis is true, $\kappa=0$, and in this case, the known space-time transformations
to the comoving system are also the transformation to the
uniformly accelerated system.

Assuming that the Clock hypothesis is not true, we show in Section 5 that
the transformations preserve a proper velocity - time interval. We predict the
existence of an \textit{unique} invariant \textit{maximal acceleration}. The proper velocity - time transformations are of Lorentz type.
We obtain an acceleration-addition
formula for relativistically admissible accelerations.
The existence of a maximal
acceleration has also been conjectured by Caianiello
\cite{Caianiello} and others.

In Section 6, we describe the W. K\"{u}ndig experiment \cite{Kundig} measuring the transverse Doppler effect.
 Kholmetski et al reanalyzed this experiment \cite{Khoimetski} and showed that in this experiment, there was a significant deviation of the time dilation predicted by Special Relativity. Their own experiment \cite{Khoimetski2} shows a similar deviation. Here we show that our model predicts an additional time
 dilation in the experiment due to the acceleration of the absorber. Based on the results of this experiment and our model,
 we predict that the Clock Hypothesis is false and that the value of the maximal
acceleration $a_m$ is of the order $10^{19}m/s^2$.

We conclude the paper with Discussion and Conclusions.
In this paper, we use SI units. Earlier results of this paper appear in \cite{FG4}.

 \section{Proper velocity and proper acceleration}

The \textit{proper velocity} ${u}$
 of an object moving with uniform velocity ${v}$ is defined by
\begin{equation}\label{tm}
{u}=\frac{{v}}{\sqrt{1-{v^2}/{c^2}}}=\gamma (v){v},
\end{equation}
where $ \gamma (v)=\frac{1}{\sqrt{1-{v^2}/{c^2}}}$. Recall that  ${u}$
is also equal to $d{r}/d\tau$, where
$d\tau=\gamma ^{-1}(v)dt$ is the proper time of the moving object. For brevity,
we will call proper velocity \textit{p-velocity}. Note that a p-velocity is expressed as a
vector of $R^3$. Conversely, any vector in $R^3$, with no
limitation on its magnitude, represents a relativistically
admissible p-velocity. The p-velocity is the spatial part of the
4-velocity.

The \textit{ proper acceleration} ${g}$ is usually defined (see \cite{Rindler}
p.71) to be the derivative of p-velocity with respect to time $t$, \textit{i.e.},
\begin{equation}\label{acceldef}
 {g}=
\frac{d{u}}{dt}\,.
\end{equation}
 Note that if an object moves with
constant proper acceleration, then its p-velocity satisfies the equation
\begin{equation}\label{accel uniform}
    \frac{d^2{u}}{dt^2}=0\,.
\end{equation}
We will say that an object is \textit{uniformly accelerated } if its proper
acceleration is constant, or equivalently, satisfies (\ref{accel uniform}). If the
 velocity of a uniformly accelerated object is parallel to the acceleration, then
it moves with the well-known hyperbolic motion (see
\cite{Moler}, \cite{Rindler} and \cite{Franklin}).

In the one-dimensional case, we have
$\frac{d{u}}{dt}=\gamma ^3\frac{d^2{r}}{dt^2}$. Moreover, the quantity $\gamma ^3\frac{d^2{r}}{dt^2}$
is invariant under Lorentz transformations between inertial systems (see \cite{Rindler} sec 3.7). Thus, in the one-dimensional case,
a uniformly accelerated motion in one inertial
system is also uniformly accelerated in any other
inertial system, implying that this property is covariant.

By a \textit{uniformly accelerated system} in this paper, we
mean a system that is uniformly accelerated with respect to a given inertial system.
Let $K$ denote an inertial system, and let $\tilde{K}$ be  a uniformly accelerated system
moving parallel to $K$ with uniform acceleration ${g}.$  For a given time $t$, we denote
by $K'$ an inertial system  which is positioned
and has the same velocity (and proper velocity) as $\tilde{K}$ at
time $t$ and moves parallel to $K$. The system
$K'$ is called a \textit{comoving system} to system $\tilde{K}$ at
time $t$.

The space-time transformation between the system $K$ and $K'$ is well known (see \cite{Moler} p.255).
If we assume the validity of the Clock hypothesis, this transformation is also the
transformation between $K$ and the uniformly accelerated system $\tilde{K}$. If we do not assume
the validity of the Clock hypothesis, it is sufficient to describe the transformation between two
comoving accelerated systems $K'$ and $\tilde{K}$, meaning that at some initial time $t_0$ their relative
velocity is zero. The inertial system $K'$ is also uniformly accelerated and its acceleration is constant and equals zero.

\section{Proper velocity - time description of events}

An important step in the derivation of the Lorentz space-time
transformations between two inertial frames is to show that such
transformations are linear. For uniformly accelerated systems, the
space-time transformation is not linear. Thus, we introduce another
description of events, called the \textit{proper velocity - time}
description, in which the transformation of events between two
uniformly accelerated systems is linear.

In the p-velocity-time description, an event is
described by the time at which the event occurred and the
p-velocity ${u}\in R^3$ of the event. The evolution of an object in a system can be
described by the p-velocity ${u}(t)$ of the object at time
$t$. The line $(t,u(t))$ replaces the world-line of special relativity in this description. To obtain the
position of the object at time $t$, we have to know the initial
position of the object and then integrate its ordinary velocity (which is readily computed from the p-velocity) with respect to time.

 To obtain the Lorentz transformations in special
 relativity, it is important that the relative position of the origins of the
frames connected with two inertial systems depends linearly on
time. This linear map expresses the relative velocity between the
systems. For uniformly accelerated systems, if we assume
that the systems are comoving at time $t=0$, the uniform acceleration
between the systems, defined by (\ref{acceldef}),
is a linear map from the time to p-velocities.

Denote by $T$ the transformation mapping
 the time and p-velocity $(t,u)$ of an event in a uniformly accelerated
system $K_g$  to the
 time and p-velocity of the same event $(t',u')$ measured in the uniformly accelerated
system $K_0$. The situation is analogous to that of the space-time transformations between two
inertial systems. In that case, the relative motion of one system with
respect to the other is described by a uniform velocity, which is a linear map from time
to space (or a line in the space-time continuum). For uniformly accelerated systems,
the relative motion one system with respect to the other is described by a uniform \textit{acceleration}, which is a linear map from time
to p-velocities (or a line in the p-velocity-time continuum). Since the space-time
transformation between two inertial systems is linear, we will assume that the p-velocity-time transformation $T$ between two uniformly accelerated systems is
also \textit{linear}.

\section{General proper velocity - time transformations between accelerated systems}

 To define the symmetry operator  between two
uniformly accelerated systems, we will use an extension of the
principle of relativity, \index{relativity!principle!general}
which we will call the \textit{General Principle of
Relativity}.\index{principle!relativity!general} This principle,
as it was formulated by M. Born (see \cite{B65}, p. 312), states
that the ``laws of physics involve only relative positions and
motions of bodies. From this it follows that no system of
reference may be favored \textit{a priori} as the inertial systems
were favored in special relativity." The principle of relativity
from special relativity states that there is no preferred
\textit{inertial} system, and, therefore, the notion of rest (zero
velocity) is a relative notion.
From the general principle of relativity, it follows that there is
no preference for inertial (zero acceleration) systems. Hence,
when considering accelerated systems, we no longer give preference
to free motion (zero force) over constant force motion. This makes
all uniformly accelerated systems equivalent.

 From the general principle of relativity, it is logical
to assume that \textit{the transformations between the
descriptions of an event in two uniformly accelerated systems
depend only on the relative motion between these systems}.
Consider now two uniformly accelerated systems
   $K_g$ and $K_0$, with a constant acceleration
  ${g}$ between them. We choose reference frames in such a way that the
description of relative motion of $K_g$ with respect to
$K_0$ coincides with the description of relative motion of
$K_0$ with respect to $K_g$. The above principle implies that the transformation $T$
mapping the description of an event in system $K_g$ to the description of the
same event in system $K_0$ will coincide with the transformation $\widetilde{T}$ from system
$K_0$ to $K_g$. This implies that $T$ is a symmetry, or $T^2=Id$.

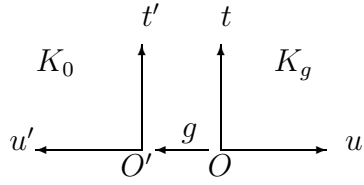
\begin{figure}[h!]
 \centering
   \begin{picture}(200,120)(-20,0)
   \put (100,50){\vector(1,0){40}}
   \put (100,50){\vector(0,1){40}}
   \put (70,50){\vector(-1,0){40}}
   \put (70,50){\vector(0,1){40}}
   \put (95,50){\vector(-1,0){20}}
   \put (147,50){$u$}
   \put (20,50){$u'$}
   \put (100,97){$t$}
   \put (70,97){$t'$}
   \put (85,55){${g}$}
   \put (120,80){$K_g$}
   \put (30,80){$K_0$}
   \put (95,40){$O$}
   \put (62,40){$O'$}
   \end{picture}
  \caption{Two uniformly accelerated systems $K_0$ and $K_g$, where system $K_0$
  moves with acceleration ${g}$ with respect to system $K_g$. The space and proper velocity axes are reversed in order to preserve the symmetry
  following from the general principle of relativity.}\label{pvelaxes}
\end{figure}

The choice of the reference frames is as follows.
  We choose the origins $O$ of $K_g$ and $O'$ of $K_0$ of the p-velocity axes to be the
  same at $t=0$, and choose the p-velocity axes  reversed, as in Figure \ref{pvelaxes}.
  We also  synchronize the clocks positioned at the origins of the frames at time
  $t=0$. Note that
  with this choice of the axes, the acceleration ${g}$ of $O'$ in $K_g$ is equal to the
  acceleration of $O$ in $K_0$, and thus the p-velocity-time transformation problem is fully
  symmetric with respect to $K_g$ and $K_0$. We will denote this
  transformation by $S_{g}$, since it is a symmetry and depends only on
  the acceleration ${g}$ between the systems.

Since the p-velocity-time
transformation $S_{g}$ is linear, it
can be represented by a $2\times 2$ matrix with
components  defined by
\begin{equation}\label{comp1a}
 \left( \begin{array}{c}  t'\\ {u'}
          \end{array} \right)=S_{g}\left( \begin{array}{c}  t\\ {u}
          \end{array} \right)= \left(
         \begin{array}{rr}
              S_{00} & S_{01} \\
              S_{10}& S_{11}
          \end{array} \right)
          \left( \begin{array}{r}  t\\ {u}
          \end{array} \right).
\end{equation}

We explain now the meaning of the  components $S_{ij}$.
The component $S_{00}$ describes the transformation of the  time $t$ in
$K_g$ of an event with p-velocity ${u}=0$ (at rest in
$K_g$) to its time $t'$ in $K_0$, and it is given by
\begin{equation}\label{tetap}
    t'=S_{00} t=\tilde{\gamma} t,
\end{equation}
for some constant $\tilde{\gamma}$. The constant $\tilde{\gamma}$ expresses the
slowdown  of the clocks  in $K_0$ due to its acceleration
relative to $K_g$. The value of $\tilde{\gamma}$ is related to the
well-known  \textit{Clock Hypothesis}. Since $K_g$ and
$K_0$ are comoving at time $t=0$, they have the same velocity at time $t=0$. Therefore, if
 the Clock Hypothesis is valid, we have $t'=t$, which implies that $\tilde{\gamma} =1$.

 To define  $S_{10}$, consider an event that
occurs at $O$, corresponding to ${u}=0$, at time $t$ in
$K_g$. Then $u'=S_{10}t$ expresses the p-velocity of this event in
$K_0$. From (\ref{tetap}), we get $u'=S_{10}\tilde{\gamma} ^{-1}t'$. Since $u'=gt'$ expresses
the relative motion of $K_g$ with respect to $K_0$, we get
\begin{equation}\label{omegap}
   S_{10}=\tilde{\gamma} g.
\end{equation}

Now we use the identity $S_{g}^2=Id$. From the matrix representation (\ref{comp1a}),
we obtain
\[S_{10}S_{00}+S_{11}S_{10}=0\;\Rightarrow\;\tilde{\gamma}^2g+S_{11}\tilde{\gamma} g=0\;\Rightarrow\;S_{11}=-\tilde{\gamma}\,.\]
We introduce a constant $\kappa$ such that $S_{01}=\tilde{\gamma}\kappa$. In this notation,
the matrix of $S_{g}$ becomes
\begin{equation}\label{matrix T alpha}
    S_{g}=\tilde{\gamma}\left(
              \begin{array}{cc}
                1 & \kappa \\
                g & -1 \\
              \end{array}
            \right).
\end{equation}
Using $S_{g}^2=Id$ once more, we get $ \tilde{\gamma}^2(1+\kappa g)=1$. Since the
time transformation preserves casuality, we get
\begin{equation}\label{alfa}
    \tilde{\gamma}=\frac{1}{\sqrt{1+\kappa g}}\,.
\end{equation}

Thus, the p-velocity time transformation between systems $K_g$ and $K_0$ is
\begin{equation}\label{coordtransprev}
  \begin{array}{cl}
    t' &=\tilde{\gamma}(t+\kappa u) \\
    u' & =\tilde{\gamma}(gt-u) .
  \end{array}
\end{equation}
Finally, by reversing the proper velocity axes in system $K_g$, we get
\begin{equation}\label{coordtranspLor}
  \begin{array}{cl}
     t' &=\tilde{\gamma}(t-\kappa u) \\
    u' & =\tilde{\gamma}(gt+u) ,
    \end{array}
\end{equation}
with $\tilde{\gamma}$ defined by (\ref{alfa}). This transformation is a\textit{ Lorentz-type transformation }.

As mentioned above, if we assume the clock hypothesis, then $\tilde{\gamma}=1$,
and, thus, from (\ref{alfa}), it follows that $\kappa=0$ in this case. Hence, if the
Clock Hypothesis is not valid, then $\kappa\neq 0$.
From now on, we will consider only the case $\kappa\neq 0$.

\section{Conservation of p-velocity-time interval and maximal acceleration}\label{sec.1.6.7}

\begin{figure}[h!]
   \begin{picture}(200,120)(-80,0)
   \put (45,10){\vector(1,0){45}}
   \put (90,10){\vector(1,0){45}}
   \put (135,10){\vector(1,0){20}}
   \put (45,00){\vector(0,1){120}}
   \put (45,10){\vector(-1,0){30}}
   \put (45,10){\line(-1,3){35}}
   \put (45,10){\line(1,1){95}}
   \put (45,10){\line(1,2){47}}
   \put (43,100){\line(1,0){4}}
   \put (15,10){\dashbox(120,90){ }}
   \put (90,10){\dashbox(00,90){ }}
   \put (47,123){$t$}
   \put (47,1){$0$}
   \put (92,3){${w}$}
   \put (13,0){$\tilde{w}$}
   \put (37,103){$1$}
   \put (89,48){${u}={g}t$}
   \put (137,3){${g}$}
   \put (20,45){$\tilde{w}$}
   \put (70,85){${w}$}
     \end{picture}
  \caption{Eigenspaces of the symmetry}\label{eigenspacep}
\end{figure}
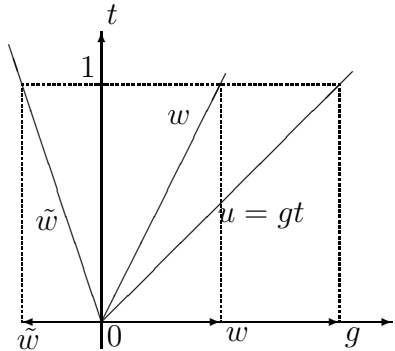

As mentioned above, the p-velocity-time transformation between the
systems $K_g$ and $K_0$ is a symmetry transformation. Such a
symmetry is a reflection with respect to the set of the fixed
points, which are the 1-eigenvectors of this transformation. We want to determine the 1-eigenvectors of $S_{g}$.
Denote by ${w}=\left(
                          \begin{array}{c}
                            w^0 \\
                            w^1 \\
                                                      \end{array}
                        \right)
$ a 1-eigenvector of $S_{g}$.  From (\ref{matrix T alpha}), it follows that this vector satisfies the system of equations
\begin{equation}\label{fixedp}
       S_{g}\left( \begin{array}{c} w^0\\ w^1
          \end{array} \right)= \tilde{\gamma}   \left(
         \begin{array}{cc}
              1 & \kappa  \\
              {g}& -1
          \end{array} \right)
         \left( \begin{array}{c} w^0\\ w^1
          \end{array} \right)= \left( \begin{array}{c}  w^0\\ w^1
          \end{array} \right)
. \end{equation}
 This system has infinitely many solutions. Thus, we may choose $w^1 =g\tilde{\gamma}$. From the second row, we have
\begin{equation}\label{1egenvect}
  w^0=1+\tilde{\gamma},\;\;w^1 =g\tilde{\gamma}\,.
\end{equation}
The meaning of this is that all the  events fixed by the
transformation $S_{g}$ are on a straight line through the
origin of the p-velocity-time continuum, corresponding to the
motion of an object with constant acceleration ${w}=\frac{w^1}{w^0}$ (see
Figure \ref{eigenspacep}) in both frames.

Similarly, for a -1-eigenvector of $\tilde{w}=\left(
                          \begin{array}{c}
                            \tilde{w}^0 \\
                            \tilde{w}^1 \\
                                                      \end{array}
                        \right)
$  of $S_{g}$, we get
 \begin{equation}\label{-1egenvect}
  \tilde{w}^0=\tilde{\gamma}-1,\;\;\tilde{w}^1 =g\tilde{\gamma}\,.
\end{equation}

We introduce a metric on the proper velocity - time continuum which makes the symmetry $S_{g}$  an isometry. Under the inner product associated with the metric, the 1 and -1
eigenvectors of $S_{g}$ will be orthogonal. The new inner
product is obtained from a metric of the form $diag(\mu,-1,-1,-1)$, where $\mu$ is an appropriate weight
for the time component with units of the square of acceleration.  The orthogonality of the eigenvectors means that
\begin{equation}\label{orth}
 < {w} | \tilde{w} >
=\mu   w^0 \tilde{w}^0- w^1 \tilde{w}^1=0.
\end{equation}
By use of (\ref{1egenvect}), (\ref{-1egenvect}) and (\ref{alfa}), this becomes
$  \mu (1+\tilde{\gamma})(\tilde{\gamma}-1)- g^2\tilde{\gamma}^2=
  -\mu\kappa g\tilde{\gamma}^2 - g^2\tilde{\gamma}^2=0$, or $\mu\kappa  + g=0$,
implying that
\begin{equation}\label{mup}
  \mu =\frac{-g}{\kappa}\quad\mbox{ and }\quad \kappa =\frac{-g}{\mu}.
\end{equation}

 From the fact that
$S_{g}$ is an isometry with respect to the inner product
with weight $\mu$, we have
\begin{equation}\label{intconservp}
  \mu (t')^2 - |{u}'|^2=\mu t^2 - |{u}|^2,
\end{equation}
which implies that our p-velocity-time
transformation from $K_g$ to $K_0$ conserves the interval
\begin{equation}\label{interval2}
ds^2=\mu dt^2 - |d{u}|^2,
\end{equation}
with $\mu$ defined by (\ref{mup}).

Note that $S_g$ maps zero interval lines in $K_g$ to zero interval lines in $K_0$. Zero interval lines correspond to motion with uniform
acceleration $\sqrt{\mu}$. Thus, for two systems $K_g$ and $K_0$
with $\kappa>0$, the acceleration $\sqrt{\mu}$ defined by (\ref{mup}) is
conserved. Obviously, the cone $ds^2>0$ is preserved
under the p-velocity-time transformation. By an argument
similar to the one in \cite{F04}, section 1.2.2, it can be shown
that $\kappa$ is independent of the relative acceleration
${g}$ between the frames $K_g$ and $K_0$. Thus, there is a
universal constant $a_m =\sqrt{\mu}$, where
$a_m$ \textbf{is the maximal acceleration.}

 Substituting $\mu=a_m^2$ into (\ref{mup}), we get
$\kappa =-g/a_m^2$. The value of $\tilde{\gamma}$ from (\ref{alfa}) becomes
 \begin{equation}\label{gama tilde}
  \tilde{\gamma}=1/\sqrt{1-g^2/a_m^2}
\end{equation}
and the proper velocity-time transformation (\ref{coordtranspLor}) becomes
\begin{equation}\label{Lorentz2}
  \begin{array}{cl}
    t' &=\tilde{\gamma}(t+gua_m^{-2}) \\
    u'_x & =\tilde{\gamma}(gt+u)\,.
    \end{array}
\end{equation}
This is a Lorentz-type transformation. Moreover,
in transformations between  accelerated systems, the interval
\begin{equation}\label{interval preservation}
   ds^2=(a_m dt)^2 - |d{u}|^2
\end{equation}
is conserved.

\section{ K\"{u}ndig's experiment and its consequences}

K\"{u}ndig's experiment (K\"{u}ndig (1963)) measured the transverse Doppler effect in a rotating disk by means of the M\"{o}ssbauer effect.
In this experiment, the distance from the center of the disk to the absorber was $R=9.3 cm$, and the rotation velocity varied between $300-35000$ $rpm$. The velocity
${v}=R\omega $ of the absorber is perpendicular to the radius, the radiation direction. K\"{u}ndig expected to measure the transverse Doppler effect by measuring
the relative energy shift, which, by relativity, should be
\begin{equation}\label{doplerShift}
  \frac{\triangle E}{E}\approx -\frac{R^2\omega ^2}{2c^2},
\end{equation}
where $E$ is the photon energy as measured from its frequency.

 Let us introduce a constant $b$ such that
 \begin{equation}\label{b def}
   \frac{\triangle E}{E}= -b\frac{R^2\omega ^2}{2c^2}.
\end{equation}
K\"{u}ndig's experimental result was
 \begin{equation}\label{Kundigres}
  b=1.0065\pm 0.011,
\end{equation}
 which was claimed to be in full agreement with the expected time dilation.

However, Kholmetskii \textit{et al} \cite{Khoimetski} found an error in the data processing of the results of K\"{u}ndig's experiment. They corrected the error and recalculated the results for three different rotation velocities for which the authors of the experiment provided all the necessary data. After their corrections,
the average value of $b$ is
  \begin{equation}\label{KholRes}
   b=1.192\pm 0.03,
  \end{equation}
 which does \textit{not} agree with (\ref{doplerShift}). They repeated a similar experiment \cite{Khoimetski2} and also observed a deviation from the usual formula for time dilation.

In \cite{F09}, it was shown that we can use the above results to show that the Clock Hypothesis is not valid. This, in turn, leads us to predict the
existence of a maximal acceleration.

The absorber is rotating. Hence, its velocity is perpendicular to the radius, and its acceleration is toward the source of radiation. Let $K$ denote the inertial frame of the lab. We can attach an accelerated system $\tilde{K}$ to the absorber. Introduce, as above, an inertial frame $K'$ comoving with the absorber. The frame $K'$ moves parallel to $K$ with constant velocity ${v}=R\omega$. The time dilation between $K$ and $K'$ is given by the transverse Doppler effect, as in (\ref{doplerShift}). If the Clock Hypothesis, claiming that there is no effect on the rate of the clock due to acceleration, is valid, then there is no change in time from system $K'$ to $\tilde{K}$. As a result, formula (\ref{doplerShift}) should
also hold for time dilation between $K$ and $\tilde{K}$. However, by (\ref{KholRes}), this is not the case, with a deviation exceeding almost 20 times the measuring error. Based on this experiment, therefore, we claim that \textbf{the Clock
hypothesis is not valid.}

In  K\"{u}ndig's experiment, the system $\tilde{K}$ moves with acceleration $a=R\omega^2$ toward the source. The transformations (\ref{Lorentz2}) are similar to the usual Lorentz transformations if we replace $v/c$ by $a/a_m$. Thus, time transformations between the inertial system $K'$ and the accelerated co-moving system $\tilde{K}$ will be given by a longitudinal Doppler type shift  by a factor $(1-a/a_m)$ due to the acceleration of $\tilde{K}$ with respect to $K'$. We have
\[  \left( 1-\frac{R\omega^2}{a_m}\right) \sqrt{1-\frac{R^2\omega^2}{c^2}}\approx\left( 1-\frac{R\omega^2}{a_m}\right)\left( 1-\frac{R^2\omega^2}{2c^2}\right)\]\[\approx1-\frac{R\omega^2}{a_m} -\frac{R^2\omega^2}{2c^2}=1-\left( 1+\frac{2c^2}{Ra_m}\right)\frac{R^2\omega^2}{2c^2}.
\]
This implies that
\begin{equation}\label{bformula}
    b=1+\frac{2c^2}{Ra_m}.
\end{equation}
Notice that the calculated value of $b$ is independent of the speed of rotation. This agrees approximately with the data \cite{Khoimetski}.

By substituting the observed time dilation in  K\"{u}ndig's experiment from (\ref{KholRes}) and $R=0.093 m$, we get
\[b=1+\frac{2c^2}{Ra_m}=1.192\pm 0.03,\]
implying that
\begin{equation}\label{maxaccelcalc}
   a_m=\frac{2c^2}{R(0.192\pm 0.03)}=(112\pm 7)c^2m^{-1}=(1.006\pm0.063)10^{19}m/s^2.
\end{equation}

\section{Discussion}

Space-time transformations between uniformly accelerated systems assuming the validity
of the Clock hypothesis  were treated
in \cite{Moler}, \cite{Rindler} among others. Within the context of conformal
transformations, they were treated by Cunningham
\cite{Gun} and Bateman \cite{Bat}, see also \cite{FEW1}. Similar transformations
appear also in Page \cite{Page1} and \cite{Page2}. As mentioned above, L. Brillouin and others argued
against the Clock Hypothesis.  For a long time, B. Mashhoon argued
against the Clock Hypothesis and developed nonlocal transformations for
accelerated observers (see the review article \cite{Mashhoon} and references therein).
Our approach treats the problem differently.

To the best of our knowledge, the transformation between uniformly accelerated systems
 described here is the only one which holds if the Clock Hypothesis is not valid.
  We have shown that the proper velocity-time transformations for such systems
 (\ref{Lorentz2}) are of Lorentz type and imply the existence of a
\textit{unique} maximal acceleration $a_m$. In this case, we predict a Doppler
shift due to the acceleration of the source in addition to its shift due to its velocity.

The existence of a maximal  acceleration for massive objects has
already been predicted by Caianiello (see  Caianiello \cite{Caianiello}, Papini and
Wood \cite{PapWu} and  Papini \textit{et al.} \cite{Pap01} and references therein).
The existence of a maximal acceleration follows also from Born's
reciprocity principle. Caianiello's model \cite{Caianiello} also supports Born's
reciprocity principle. From Caianiello's model, the estimate of the maximal
acceleration in Scarpetta \cite{scarpetta84} is $a_m=5\cdot10^{50}g.$
We are not aware of any previous derivation of the maximal acceleration
from the non-validity of the Clock Hypothesis.

The W. K\"{u}ndig experiment \cite{Kundig}, as reanalyzed  by Kholmetski et al \cite{Khoimetski}, is the first experiment showing that the Clock Hypothesis is not valid.
It predicts that the value of the maximal acceleration $a_m$ is of the order $10^{19}m/s^2$.

The Clock Hypothesis was tested in the Muon Storage Ring experiment of J. Bailey \textit{et al.}  \cite{Baily77} where
they claimed ``no effects on the particle lifetime are seen in this experiment where the transverse acceleration is
 $\sim 10^{18}g$." In the experiment, the muons were rotating on a ring of radius $R=7m$. The transverse proper acceleration in the experiment was $a= \gamma c^2/R\approx 3.77\cdot10^{17}m/s^2$, which, by (\ref{gama tilde}), gives a time-dilation correction due to acceleration of order $a^2/(2a_m^2)\approx 7\cdot 10^{-4}$. This is significantly less than the accuracy of the experiment. Thus, this experiment does not contradict our model.

 The novel experimental laser research based on the Sagnac effect improved significantly the accuracy with which
 non-inertial effects are measured (see \cite{VetStedman} and \cite{Malykin}). Hence, one would expect to observe in these experiments deviations from
 Special Relativity, as in (\ref{KholRes}). However, to the best of our knowledge, no such deviation was observed. The reason for this
 is as follows. The deviation of $b$ from 1 in K\"{u}ndig's experiment was caused by the relative acceleration of the source and the absorber. This acceleration caused a correction in time dilation of the order \textit{one} in $a/a_m$. In the rotating ring experiments, however, there is \textit{no} acceleration between the source and the detector. Thus, the time dilation correction due to the acceleration is of the order \textit{two} in  $a/a_m$, which is hard to detect.


\section{Conclusion}
In this paper we give a description of transformations (\ref{Lorentz2}) between accelerated systems without the Clock Hypothesis. We established the connection of this hypothesis to the maximal acceleration. We predict a Doppler shift due to the acceleration of the source. Based on this, we give a theoretical explanation of the time dilation deviation from SR in two experiments and give a first experimental estimate of the maximal acceleration.

K\"{u}ndig's experiment was not designed to test the maximal acceleration. Thus, it is only an indication of the existence of a maximal acceleration and an estimate of its value. On the other hand, an experiment determining the value of maximal acceleration could be done with currently available technology.

\ack

We would like to thank  A. Gavrilov and the referees for important remarks and suggestions, and T. Scarr for editorial comments.

\end{document}